\documentclass[twocolumn,english,aps,prb,floatfix]{revtex4}
\usepackage[T1]{fontenc}
\usepackage[latin9]{inputenc}
\usepackage{textcomp}
\usepackage{wasysym}
\usepackage{graphicx}

\makeatletter

\providecommand{\tabularnewline}{\\}

\@ifundefined{textcolor}{}
{%
 \definecolor{BLACK}{gray}{0}
 \definecolor{WHITE}{gray}{1}
 \definecolor{RED}{rgb}{1,0,0}
 \definecolor{GREEN}{rgb}{0,1,0}
 \definecolor{BLUE}{rgb}{0,0,1}
 \definecolor{CYAN}{cmyk}{1,0,0,0}
 \definecolor{MAGENTA}{cmyk}{0,1,0,0}
 \definecolor{YELLOW}{cmyk}{0,0,1,0}
 }

\makeatother

\usepackage{babel}

\begin{document}

\preprint{This line only printed with preprint option}

\title{The Absence of Vortex Lattice Melting in a Conventional Superconductor}

\author{C. J. Bowell,$^{1*}$ R. J. Lycett,$^{1}$ M. Laver,$^{2}$ C. D.
Dewhurst,$^{3}$ R. Cubitt,$^{3}$ E. M. Forgan$^{1}$ }

\affiliation{$^{1}$School of Physics and Astronomy, The University of Birmingham,
Birmingham, B15 2TT, UK}

\affiliation{$^{2}$Laboratory for Neutron Scattering, Paul Scherrer Institut,
5232 Villigen PSI, Switzerland }

\affiliation{$^{3}$Institut Laue Langevin, BP 156, F-38042 Grenoble, France}
\begin{abstract}
The state of the vortex lattice extremely close to the superconducting
to normal transition in an applied magnetic field is investigated
in high purity niobium. We observe that thermal fluctuations of the
order parameter broaden the superconducting to normal transition into
a crossover but no sign of a first order vortex lattice melting transition
is detected in measurements of the heat capacity or the small angle
neutron scattering (SANS) intensity. Direct observation of the vortices
via SANS always finds a well ordered vortex lattice. The fluctuation
broadening is considered in terms of the Lowest Landau Level theory
of critical fluctuations and scaling is found to occur over a large
$H_{c2}(T)$ range. 
\end{abstract}
\maketitle

\section{Introduction}

Thermal fluctuations are large in the high-\emph{T}$_{c}$ superconductors
and their signature can be seen over a significant region of the superconducting
phase diagram. In high quality samples, the effects of disorder can
be negligible compared with thermal fluctuations and a clear first
order transition, associated with melting of the vortex lattice \cite{key-1,key-2}
occurs well below the \textquotedblleft{}mean-field\textquotedblright{}
superconducting to normal transition temperature, \emph{T}$_{c2}$(\emph{H}).
However, in a low-\emph{T}$_{c}$ sample, thermal fluctuations are
significantly smaller due to the low temperature, the high vortex
rigidity and strong inter-vortex forces. Using the Lindemann criterion
(that the lattice melts when fluctuations reach a critical fraction
of the lattice spacing) vortex lattice melting is predicted \cite{key-3}
to occur only several millikelvin below \emph{T}$_{c2}$(\emph{H}). 

The small thermal energy in low-\emph{T}$_{c}$ superconductors makes
pinning of the vortex lattice much more significant \cite{key-4},
hence phase transitions related to this pinning become an alternative
theoretical scenario \cite{key-5,key-6}. For example, a disordering
transition in the vortex lattice of niobium has been observed \cite{key-7}
directly by small angle neutron scattering (SANS) and is associated
with the peak effect \cite{key-8} (increased pinning of the vortices,
linked to a softening of the vortex lattice below \emph{T}$_{c2}$(\emph{H})).
The position of this transition varies with niobium sample purity
\cite{key-9} and is not observed, to within 20 mK of \emph{T}$_{c2}$(\emph{H}),
in a high purity sample \cite{key-10}. Heat capacity measurements
on Nb$_{3}$Sn \cite{key-11} show signs of a vortex lattice phase
transition in a sample with significant vortex lattice pinning (and
showing the peak effect), hence it is not clear whether this transition
is thermal melting or pinning-influenced disordering. To investigate
thermal melting of the vortex lattice, a high-quality low-pinning
sample is clearly important. Here we present heat capacity and SANS
measurements on high purity niobium. We observe the effects of thermal
fluctuations of the superconducting order parameter, but find no evidence
for either a first order melting transition or for order-disorder
transitions associated with vortex pinning.

\section{Methods}

Heat capacity measurements were performed using the a.c. technique
of Sullivan and Seidel \cite{key-13}. The sample is strongly thermally
connected to a thermometer and heater, and weakly connected to a bath.
An a.c. current of frequency $\frac{1}{2}\omega$ is passed through
the heater, so that heat is generated at a rate $\frac{1}{2}P_{0}\left[cos(\omega t)+1\right]$.
Conditions are such that: a) The heat capacities of the thermometer
and heater are much less than that of the sample. b) The sample heater
and thermometer come into equilibrium with a time constant much less
than the inverse of the angular frequency, $\omega$. c) The sample
to bath relaxation time is much longer than the inverse frequency.
Under these conditions the temperature of the sample is given by a
d.c. increase above the temperature of the bath, plus an a.c. component,
\[
T_{a.c.}=P_{0}cos(\omega t)/(2\omega C),\]
where \emph{C} is the heat capacity of the sample (plus negligible
contributions from heater and thermometer). The heat capacity can
then be measured by accurate measurement of the size of the a.c. temperature
oscillation. One important caveat of using this technique is that
changes in the heat capacity over a range equal to the amplitude of
the temperature oscillation are smoothed over. It is therefore important
to set the size of the temperature oscillation to be small enough
that sufficiently detailed changes in the heat capacity can be resolved.
The size of the a.c. temperature oscillation at the sample was therefore
set at less than 1 mK (typically 0.2 mK).

\begin{figure}

\includegraphics[bb=0bp 100bp 720bp 400bp,clip,width=8.5cm]{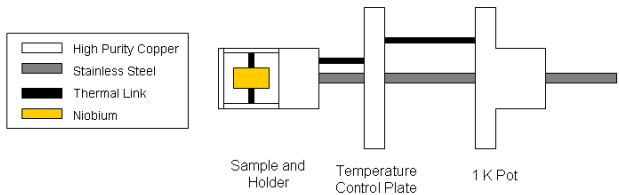}

\caption{(color online) Schematic of the cryostat insert for the heat capacity
measurements showing how the sample holder connects to the 1 K Pot
via the temperature control plate. All of these were mounted inside
a vacuum can, with no exchange gas. Heaters and thermometers were
present on the sample, on the temperature control plate and on the
1 K Pot. \label{fig:Schematic-of-the}}

\end{figure}

A schematic of the cryostat insert is shown in Figure \ref{fig:Schematic-of-the}.
A 14 g niobium sample (Nb1) was attached to a copper holder by nylon
thread, and thermally connected to it by copper wire to give a thermal
time constant of $\sim$ 3 s to the holder. Heat was applied to the
sample by passing a current at 4 Hz along a small amount of resistive
wire wound non-inductively around the sample. Measurement of the oscillatory
part of the sample temperature was performed using a ground-down GaAlAs
diode; the a.c. component of the voltage across it was measured using
a lock-in-amplifier (LIA) at 8 Hz. The sample holder consisted of
a large block of copper to act as a low-pass thermal filter. The sample
holder was strongly thermally connected to a temperature control plate,
where the base temperature of the sample was controlled above that
of the 1 K Pot during a measurement scan. The required temperature
stability ($\sim$ 20 $\mu$K) was achieved using a home-made analog
PID (proportional, integral, derivative) feedback system that heated
the plate. The sensor was a high-sensitivity carbon resistance thermometer
attached to the temperature control plate, with its resistance measured
by an audio-frequency a.c. bridge with a LIA as null detector. Fine
temperature scans were made by adding a ramped d.c. voltage to the
LIA d.c. output to shift the control point of the PID controller.
Absolute measurement of the temperature at the temperature control
plate was achieved by a calibrated Ge thermometer mounted onto the
plate (but away from the applied magnetic field). The sample temperature
was corrected to take into account the d.c. component of the heating
at the sample. 

During a measurement, the temperature was ramped with very small steps
through the superconducting transition at fixed field both on warming
and on cooling. Normal state measurements were taken 20 mT higher
in field. A typical scan took 24 hours. A measurement of the amplitude
of the a.c. voltage from the GaAlAs thermometer was typically taken
every 300 s, with the lock-in-amplifier averaging the signal with
a time constant of typically 100 s. This averaging introduced a lag
between warming and cooling measurements which has been corrected
for in the measurements presented in the paper.

The sample was aligned in the centre of a magnetic field with uniformity
3 parts in 10$^{4}$ over the volume of the sample. Two small superconducting
coils were arranged either side of the sample to allow an optional
a.c. field of amplitude 0.7 G to be applied perpendicular to the main
field. 

Small angle neutron scattering (SANS) measurements were performed
on the instrument D22 at the Institut Laue Langevin, Grenoble but
with the instrument temperature control enhanced to achieve the required
ultra-fine temperature control. While the external electronics were
the same as the heat capacity measurements, the nature of neutron
cryostats required a different set-up in the cryostat insert. There
was a low pressure of helium exchange gas in the sample chamber and
its temperature was set by the relatively coarse temperature control
of the instrument. The sample was mounted in a copper can inside the
sample chamber; the can temperature was raised above that of the instrument
control using a heater and carbon resistance thermometer attached
inside the can, and PID feedback was used, as in the heat capacity
measurements to perform temperature sweeps. The magnet that provided
the main magnetic field was designed to give uniformity 1.5 x 10$^{-4}$
over a 24 mm diameter spherical volume. As with the heat capacity
measurements, additional coils allowed a small a.c. field of amplitude
0.1 G to be applied to the sample. 

Our experiments required samples in which vortex lattice pinning was
minimal. The 14 g, barrel shaped, niobium sample (residual resistance
ratio, RRR = 1000) was made high-purity and homogeneous by annealing
in ultra-high vacuum (< 3 $\times$ 10$^{-9}$ torr) for 48 hr at
2300 K. It was then oxygen treated \cite{key-14} to reduce surface
pinning. This sample was used in both the heat capacity measurements
and SANS measurements, and is known as Nb1. Two other samples (used
in the SANS measurements), Nb2 and Nb3, were spherical, 11 g (RRR
= 450) and ellipsoidal, 8 g (RRR = 1000) respectively, and prepared
in the same way.

\section{Heat Capacity Measurements}

\begin{figure*}
\includegraphics[clip,width=12cm]{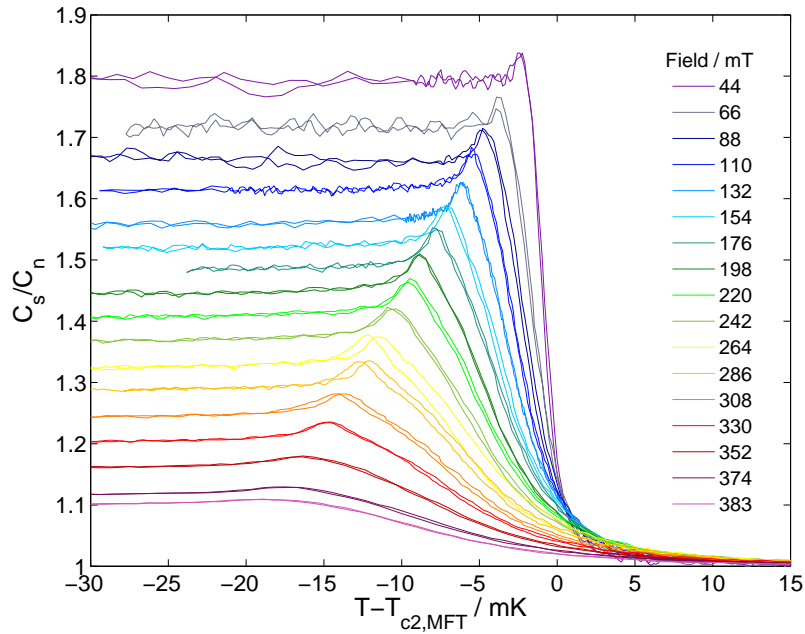}

\caption{(color online) The ratio of the superconducting and normal state heat
capacities, \emph{C}$_{s}$/\emph{C}$_{n}$, measured as a function
of temperature in the niobium sample Nb1. The magnetic field was applied
parallel to the {[}111{]} crystal axis. Normal state measurements,
$C_{n}$, were taken 20 mT higher in field. The temperature has been
plotted as \emph{T}-\emph{T}$_{c2,MFT}$, where \emph{T}$_{c2,MFT}$(\emph{H})
is the value the transition temperature would have in the absence
of fluctuations and is taken as the temperature at which the fluctuation
heat capacity is a fraction, 0.2, of the mean-field jump in \emph{C}
\cite{key-18}. When a small hysteresis is seen, measurements taken
on warming show the higher transition temperature.\label{fig:C1}}

\end{figure*}
\begin{figure*}[t]
\begin{tabular}{cccc}
\textbf{a} & \includegraphics[angle=-90,width=9cm]{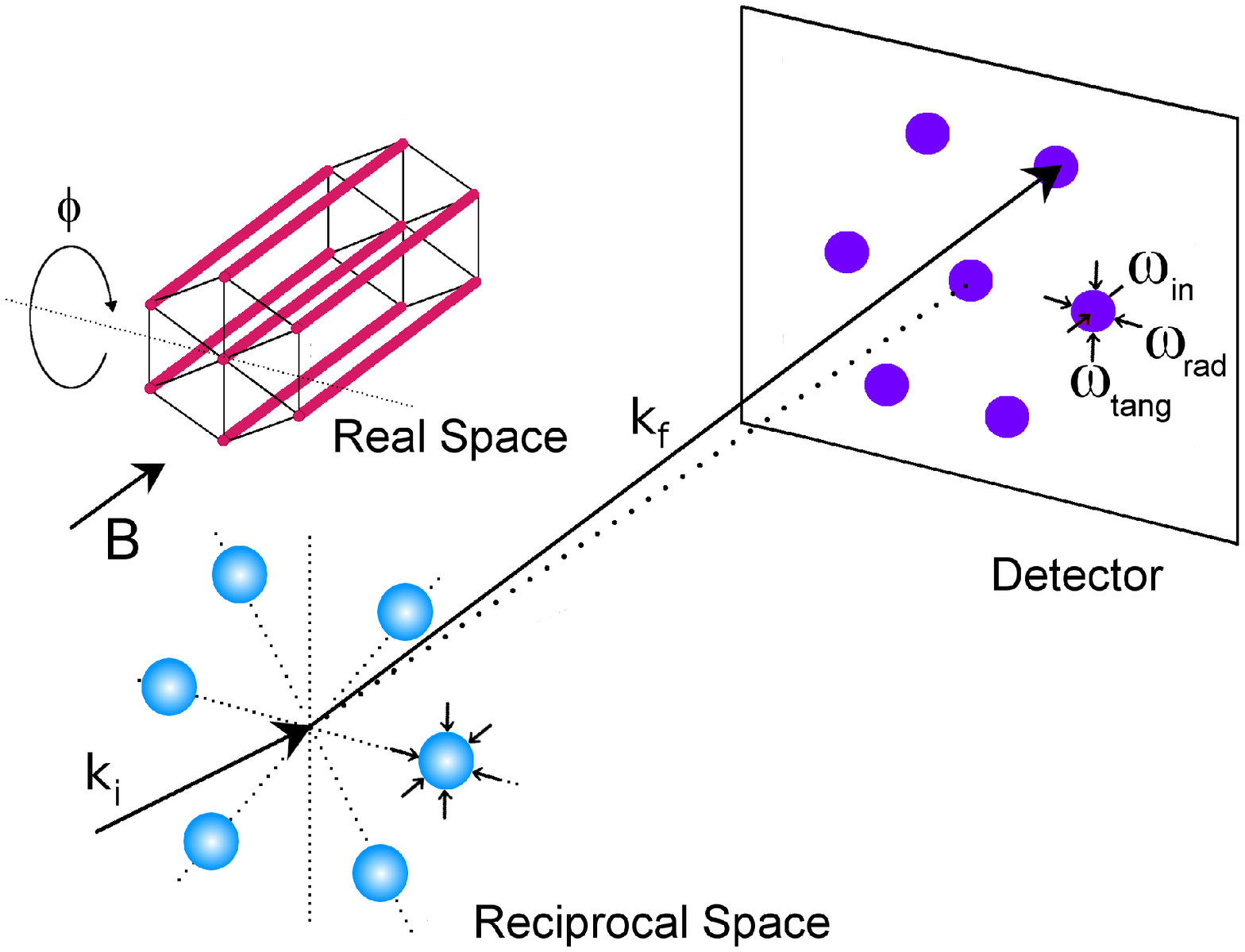} & \textbf{b} & \includegraphics[angle=-90,width=6.5cm]{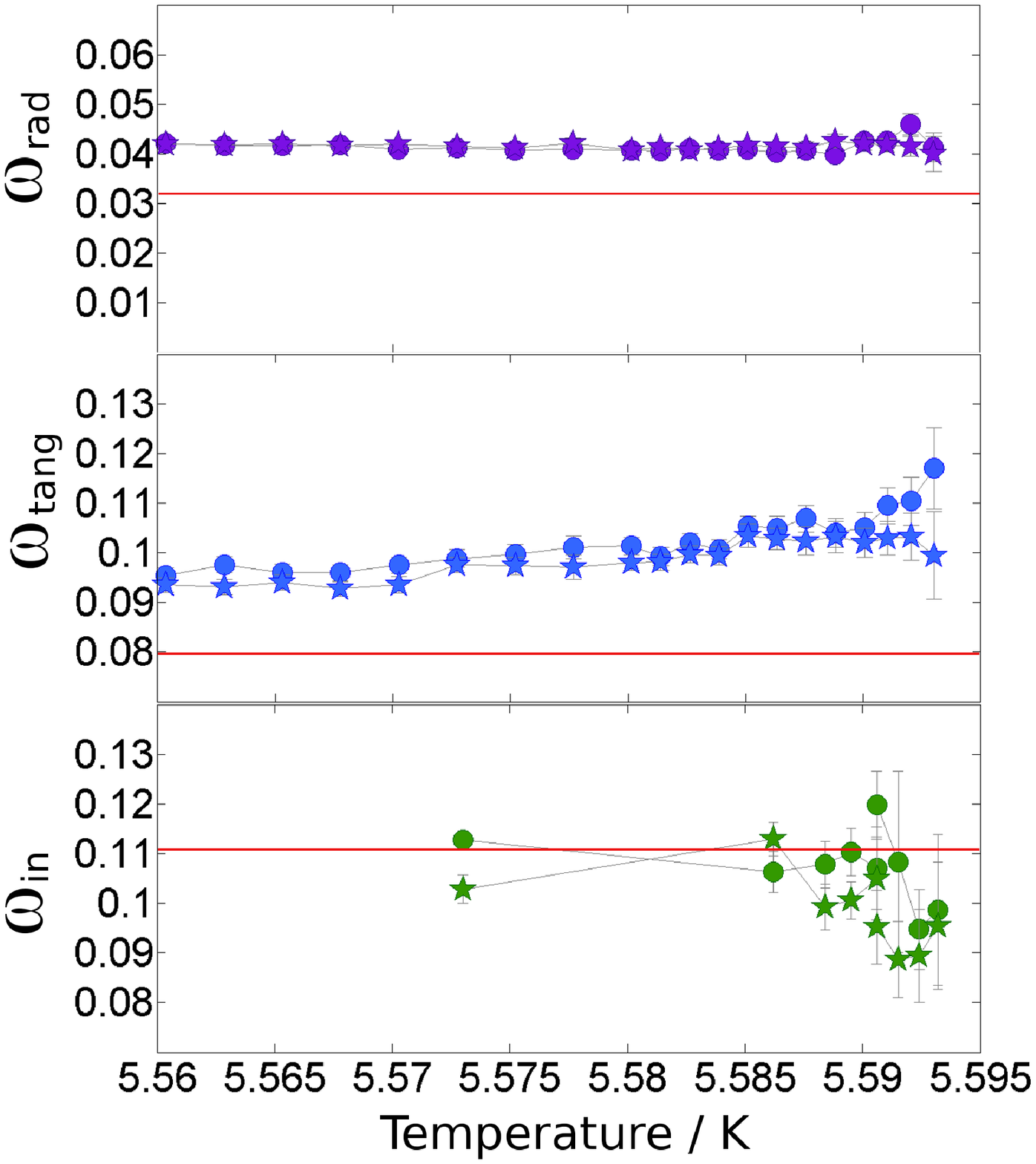}\tabularnewline
\end{tabular}

\caption{(color online) (a) Schematic to illustrate the geometry of a SANS
experiment, where \textbf{k}$_{i}$ and \textbf{k}$_{f}$ represent
the incident and scattered neutron beams respectively. (b) The FWHM
of Gaussian fits to the size of diffraction spots at the detector
in the temperature regime of critical fluctuations at 200 mT (applied
parallel to the {[}111{]} axis of Nb1). Measurements were taken on
cooling and are shown for all temperatures at which there was sufficient
diffracted intensity to fit the spot shape. Backgrounds taken at 5.610
K were subtracted before fitting. For measurements of $\omega_{rad}$
and $\omega_{tang}$, $\phi$ was fixed at the angle that brought
the top two first order spots from the hexagonal lattice, left (\CIRCLE{})
and right ($\star$), simultaneously onto the Bragg diffraction condition
and maximised their intensity. Red lines mark the FWHM expected for
a perfect lattice, where the width is due solely to instrument resolution.
\label{fig:SANS1}}

\end{figure*}

Measurements of the heat capacity of niobium were the first to show
that critical fluctuations broaden \emph{T}$_{c2}$(\emph{H}) from
a second order phase transition into a crossover \cite{key-12}. While
no first order vortex lattice melting transition was observed in these
measurements, they were performed at constant magnetic induction \emph{B},
whereas constant magnetic field \emph{H} is required to observe a
first order transition clearly. Figure \ref{fig:C1} presents our
measurements of the heat capacity of high purity niobium over the
full \emph{H}$_{c2}$(\emph{T}) range. The surface treatment given
helps to make the sample almost perfectly magnetically reversible,
to allow the measurements to be performed at constant \emph{H}. However,
a small (< 0.5 mK) hysteresis can be seen at intermediate fields,
resulting from some residual surface pinning of vortices and corresponding
to flux trapping of less than 0.1 mT. Magnetisation measurements on
a smaller sample, similarly treated, showed no sign of the peak effect. 

As shown in Figure \ref{fig:C1}, we observe a transition broadened
by critical fluctuations over the range $\sim$ 5 to 30 mK (depending
on field), giving a peak below the mean field transition and a long
tail above it. At no temperature do we see sign of a first order phase
transition, of the type observed in the high-\emph{T}$_{c}$ superconductors,
which would appear as a spike in the heat capacity. In YBa$_{2}$Cu$_{3}$O$_{7-\delta}$
the entropy of vortex lattice melting was found to be 0.4 k$_{B}$
per vortex per superconducting layer \cite{key-2}. In comparing the
latent heat in a layered system with that in isotropic niobium, we
take the layer spacing to be equivalent to the coherence length, $\xi$.
Using parameters suitable for our niobium sample we estimate (Appendix)
that melting would cause a spike in the heat capacity of height $\Delta$C$_{melt}$
$\sim$ 1 mJ mol$^{-1}$ K$^{-1}$ in a field of 300 mT (for a transition
width $\Delta$\emph{T}$_{melt}$ $\sim$ 2 mK); this is $\sim$ 10
\% of the mean field jump, $\Delta C$, at the superconducting to
normal transition, which would be readily resolvable in our measurements
but is clearly not seen.

\section{SANS Measurements}

\begin{figure*}
\includegraphics[width=12cm]{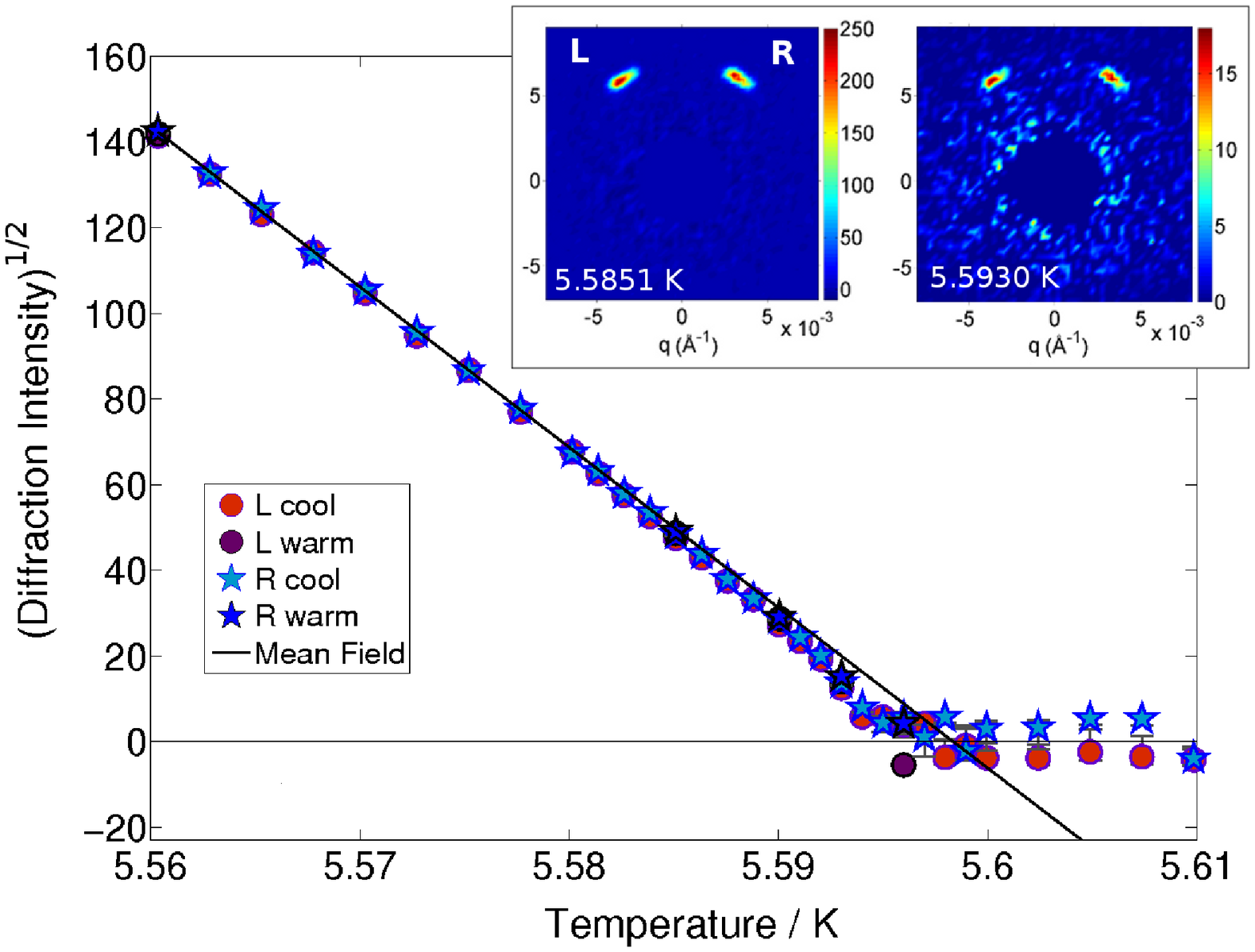}

\caption{(color online) The square root of the diffracted intensity minus background,
$\sqrt{I}$~(~$\propto$~|\emph{F}$_{hk}$|), as a function of
temperature in a field of 200 mT, measured during the same scan that
$\omega_{rad}$ and $\omega_{tang}$ (Figure 2(b)) were recorded.
Negative values of \emph{$\sqrt{I}$} (calculated as sign(\emph{I})$\sqrt{|I|}$)
result from the background subtraction. The mean field theory linear
behaviour is obtained from a fit to the temperature dependence outside
the critical region. Insets show the diffraction pattern observed
on the detector (with direct beam masked and background subtracted)
for two temperature points. \label{fig:Figure sansintensity}}

\end{figure*}

We can also search for a melting transition by examining the state
of the vortex lattice using SANS. This technique was used in the observation
of melting/decoupling of \textquoteleft{}pancake\textquoteright{}
vortices in the high-\emph{T}$_{c}$ material BSCCO \cite{key-15}
but the low scattering intensity allowed no investigation of the nature
of the disordered vortex state. The straightforward geometry of SANS
allows the size of the diffraction spot on the detector to be related
to correlation lengths of the vortex lattice \cite{key-16}. This
idea is illustrated in Figure \ref{fig:SANS1}(a). The width $\omega_{in}$
is obtained from a {}``rocking curve'' of diffraction intensity
as a function of angle when $\phi$, the angle between the applied
magnetic field/sample {[}111{]} axis and the incident neutron beam,
is rocked through the Bragg condition. On the peak of the rocking
curve, $\omega_{rad}$ and $\omega_{tang}$ are the widths of the
diffraction spot on the detector, parallel and perpendicular respectively
to the relevant \emph{q}-vector. The collimation and wavelength spread
contribute to these widths, but there are additional contributions
due to the crystallinity of the vortex lattice in the sample. Mosaic
spread (i.e. meandering of the average vortex line direction) is mainly
seen in $\omega_{in}$. Any spread in lattice spacing is more apparent
in $\omega_{rad}$, while the spread in vortex lattice orientation
(corresponding to rotations of the vortex lattice about the field
direction) contributes to $\omega_{tang}$. 

Measurements of these widths, on the same sample used in the heat
capacity measurements, Nb1, are presented in Figure \ref{fig:SANS1}(b).
The perfection of the vortex lattice is such that the width of the
rocking curve, $\omega_{in}$, is limited by the instrument resolution
while no changes in the radial spot width $\omega_{rad}$ are observed.
There is a small increase in $\omega_{tang}$, the tangential width
of the spot, but the $\sim$ 0.01\textdegree{} change is extremely
small compared with the spread that would be observed for an orientationally
disordered vortex lattice, which would give a ring of diffracted intensity.

A signature of melting would also appear in the intensity, \emph{I}$_{hk}$,
of a diffraction spot, integrated over the rocking curve. This depends
on the vortex lattice form factor, \emph{F}$_{hk}$, which expresses
the {}``field contrast'' in the mixed state,

\[
I_{hk}=2\pi\mu V\left(\frac{\gamma}{4}\right)^{2}\frac{\lambda_{n}^{2}}{\Phi_{0}^{2}q_{hk}}\left|F_{hk}\right|^{2},\]

where $\mu$ is the incident neutron flux, $V$ is the sample volume,
$\gamma$ is the neutron magnetic moment in nuclear magnetons, $\lambda_{n}$
is the neutron wavelength and $\Phi_{0}$ is the flux quantum. Mean-field
Ginzburg-Landau theory \cite{key-17} gives \emph{F}$_{hk}$ proportional
to the magnetisation, \emph{M}, \[
F_{hk}=(-1)^{v}e^{-\pi v/\sqrt{3}}\mu_{0}M,\]
 where $v=h^{2}+hk+k^{2}.$ In the absence of fluctuations the magnetisation,
and hence the form factor, should go linearly to zero as \emph{T}
$\rightarrow$ \emph{T}$_{c2}$(\emph{H}). This linear variation of
form factor with temperature has been verified to within 20 mK of
$T_{c2}(H)$ on a high purity niobium sample \cite{key-10}, however
that experiment did not have sufficient temperature resolution to
probe the region where vortex lattice melting would be expected to
occur. Measurement of the integrated intensity this close to \emph{T}$_{c2}$(\emph{H})
is only possible due to the relatively large field contrast in the
mixed state of niobium. The equivalent experiment on a superconductor
with a longer penetration depth would require unfeasibly long count
times \cite{key-15}. 

As shown in Figure 2(b), the rocking curve width, $\omega_{in}$,
for the vortex lattice in our sample is essentially independent of
temperature in the critical region. Hence, apart from a proportionality
constant, the temperature variation of $\left|F_{hk}\right|$ may
be obtained by tilting the sample to the peak of the rocking curve
and measuring diffraction intensity versus temperature with $\phi$
fixed. In the critical fluctuation region, we observe (Figure \ref{fig:Figure sansintensity}
) that the square-root of the diffracted intensity, $\sqrt{I}$, falls
below the linear mean-field theory prediction, but even in this region
the lattice remains well ordered (inset Figure \ref{fig:Figure sansintensity}).
A first order melting transition would appear as a discontinuity in
the intensity, and this is not observed. It might be argued that temperature,
field or sample inhomogeneity could provide the smearing of the SANS
results. However, if we suppose that the measurements represent a
smearing of mean field linear variation of $\sqrt{I}$ with temperature
near \emph{T}$_{c2}$ the intensity would lie \emph{above} the mean
field extrapolation from lower temperatures, and not below it as we
observe.

\section{Critical Scaling}

\begin{figure*}
\begin{tabular}{cccc}
\textbf{a} & \includegraphics[bb=0bp 120bp 600bp 700bp,clip,width=8.5cm]{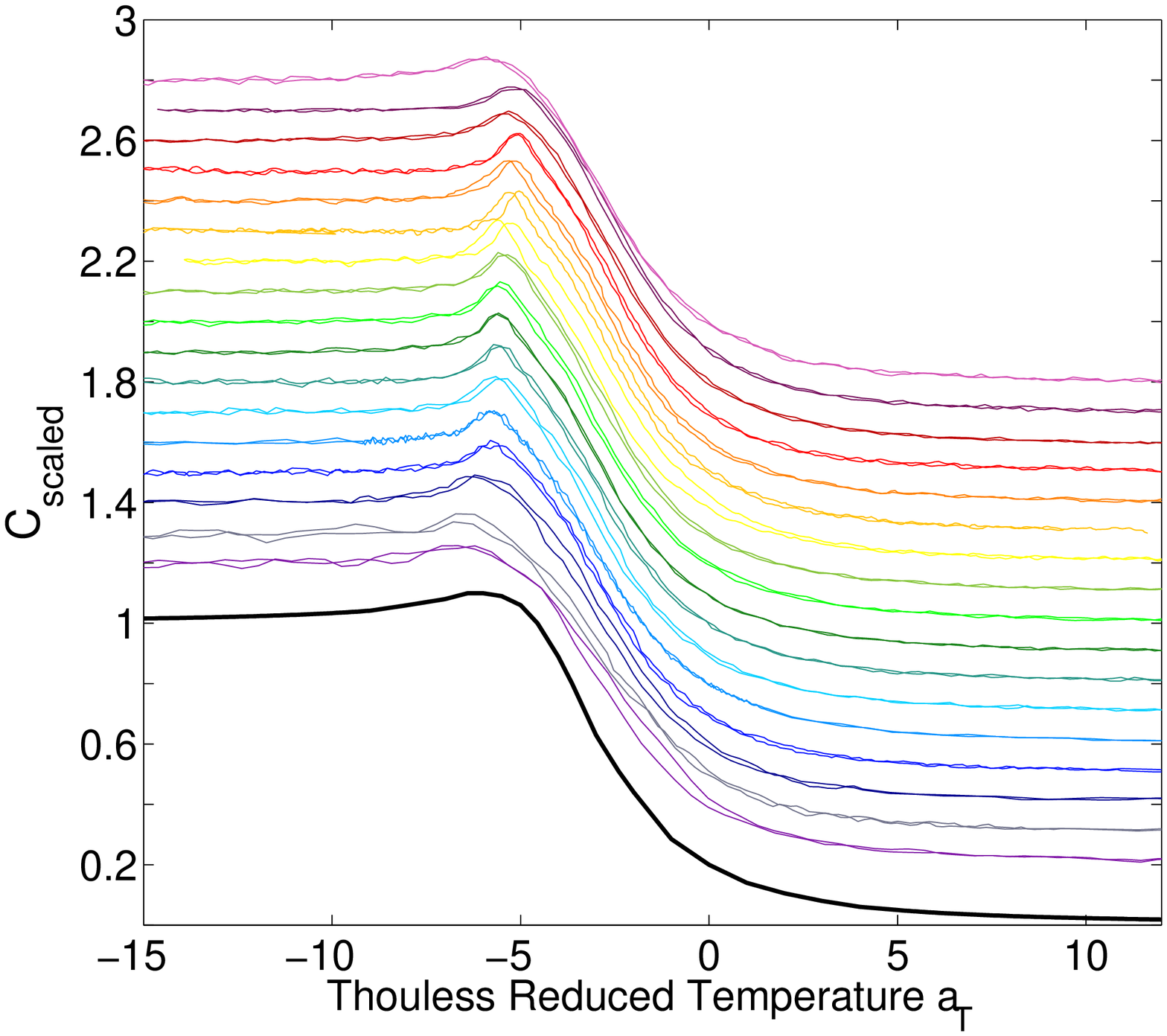} & \textbf{b} & \includegraphics[bb=0bp 120bp 600bp 700bp,clip,width=8.5cm]{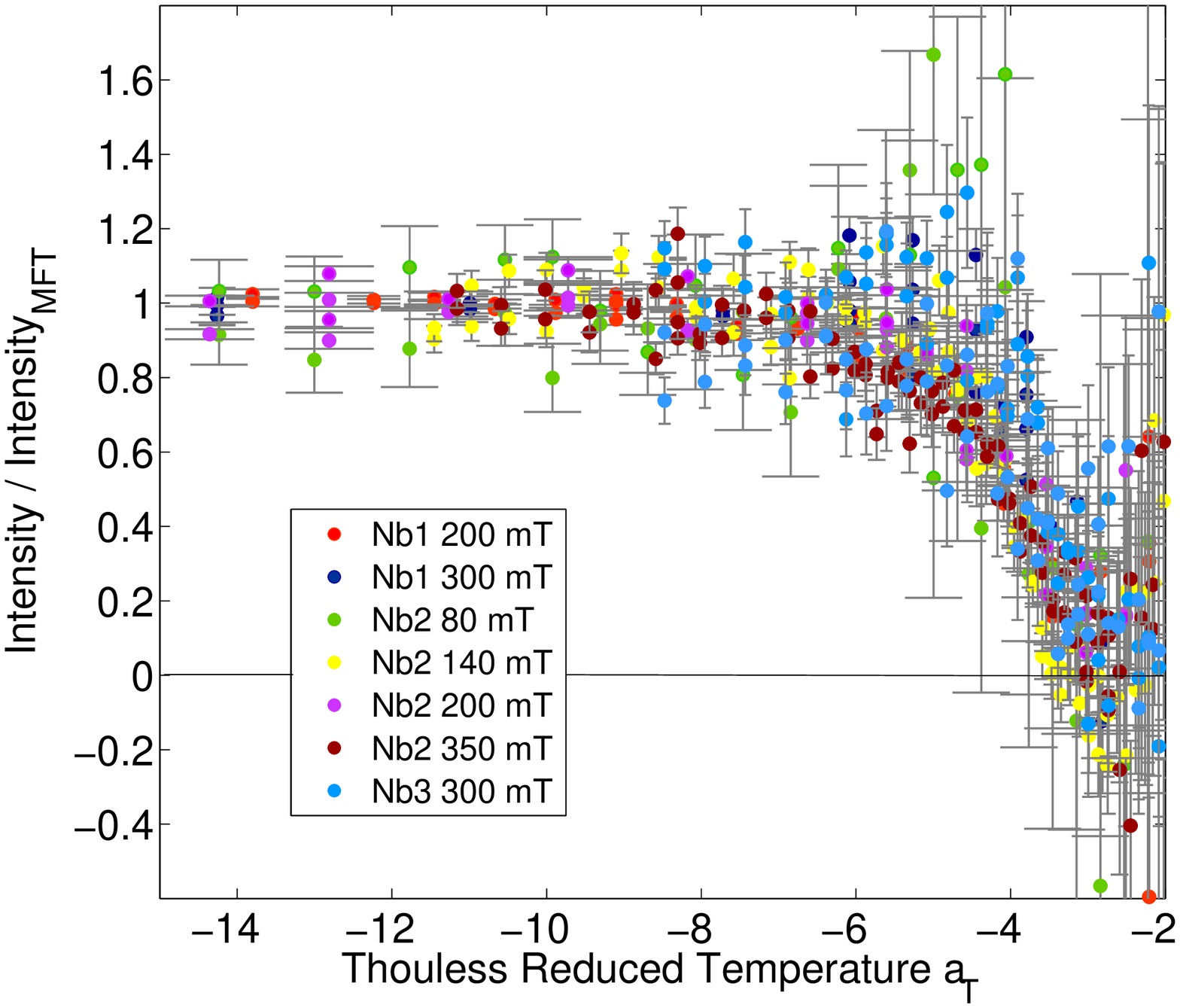}\tabularnewline
\end{tabular}

\caption{(color online) Scaling of the heat capacity and SANS intensity. (a)
The heat capacity measurements of Figure 1, normalised to the mean-field
behaviour as (\emph{C}$_{s}$~-~\emph{C}$_{n}$)~/~(\emph{C}$_{s,MFT}$~-~\emph{C}$_{n}$),
plotted on the Thouless reduced temperature scale, a$_{T}$. \emph{C}$_{s,MFT}$
was extrapolated through the transition from a linear fit of \emph{C}$_{s}$/\emph{C}$_{n}$
versus temperature below the fluctuation region \cite{key-12}. The
measurement at 44 mT is shifted vertically by 0.2 units, to avoid
overlap with the Thouless theoretical prediction \cite{key-18}, which
is plotted as the black line. All successive plots (at the same sequence
of fields as in Figure 1) are shifted by 0.1 units. (b) SANS intensity,
normalised as \emph{I} / \emph{I}$_{MFT}$, where \emph{I}$_{MFT}$
was found from a linear fit of $\sqrt{I}$ versus temperature below
the critical region (as shown for example in Figure 2) plotted versus
the Thouless reduced temperature a$_{T}$. Measurements are shown
for three samples, Nb1: barrel shaped, Nb2: spherical and Nb3: ellipsoidal.
All samples were prepared under similar condition with similar \emph{RRR}
values. As no significant difference was observed between the left
and right hand diffraction spots, or on warming and cooling, these
measurements have been plotted with the same marker style for clarity.
The error bars become large at high temperatures because the data
points represent the ratio of two quantities, both of which are tending
to zero. \label{fig:fluct}}

\end{figure*}

To examine the origin of these deviations from mean field theory we
look for a universal temperature scale which would allow measurements
at different fields to collapse onto the same curve. 3D XY scaling
is not suitable, as it requires a spatially uniform order parameter,
which is not valid in niobium as the vortex cores are large. However,
the Lowest-Landau-Level Ginzburg-Landau (LLL-GL) theory of critical
fluctuations is appropriate. The relevant temperature scale $a{}_{T}$,
as calculated by Thouless \cite{key-18}, is given by $a_{T}=(T-T_{c2,MFT})/\delta$,
where

\[
\delta=\left(\frac{k_{B}}{8\pi\xi_{0}^{3}\Delta C}\right)^{\frac{3}{2}}\left|\left(\frac{T}{B_{c2}(0)}\right)\frac{dB_{c2}(T)}{dT}\right|^{\frac{1}{3}}\left(\frac{B}{B_{c2}(0)}\right)^{\frac{2}{3}}T.\]
Critical scaling of the heat capacity measurements, Figure \ref{fig:fluct}(a),
is successful over a large region of the phase diagram, except at
the highest and lowest fields. This confirms the fluctuation origin
of the signal and rules out inhomogeneity broadening. In the field
region where our measurements scale, we find that the long tail in
heat capacity at high temperatures agrees well with the Thouless prediction
within LLL-GL, but the enhanced peak is narrower than expected from
theory. This may demonstrate the need for higher-order terms in the
fluctuations to be included in the theory, mixing of Landau levels
at the transition \cite{key-19}, or a consequence of coupling between
the vortex and crystal lattice (which also breaks the rotational degeneracy
of the vortex lattice and gives it a preferred orientation). 

\begin{figure}
\begin{tabular}{cc}
\textbf{a} & \includegraphics[bb=60bp 50bp 520bp 700bp,clip,angle=270,width=8.5cm]{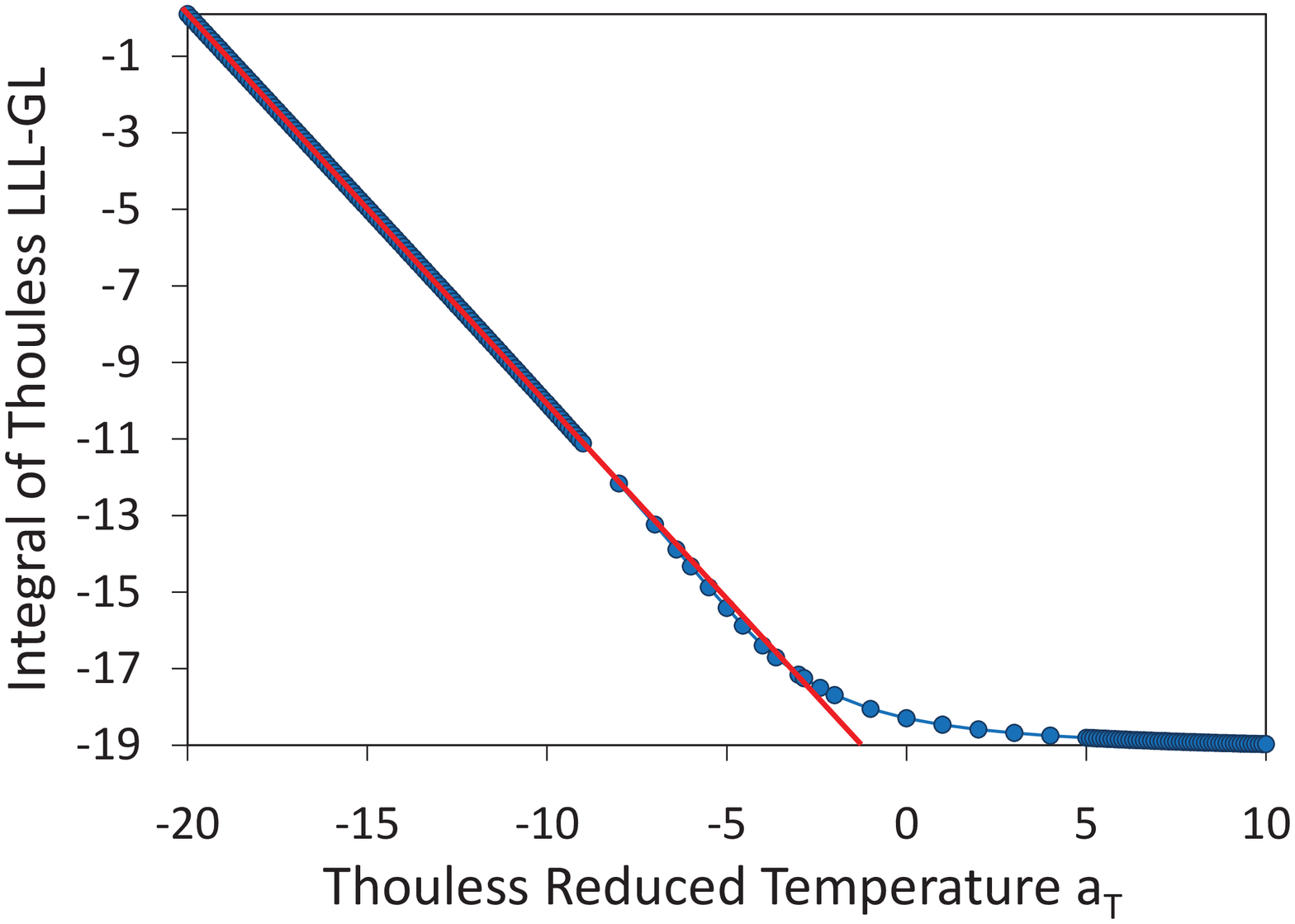}\tabularnewline
\textbf{b} & \includegraphics[bb=60bp 50bp 520bp 700bp,clip,angle=270,width=8.5cm]{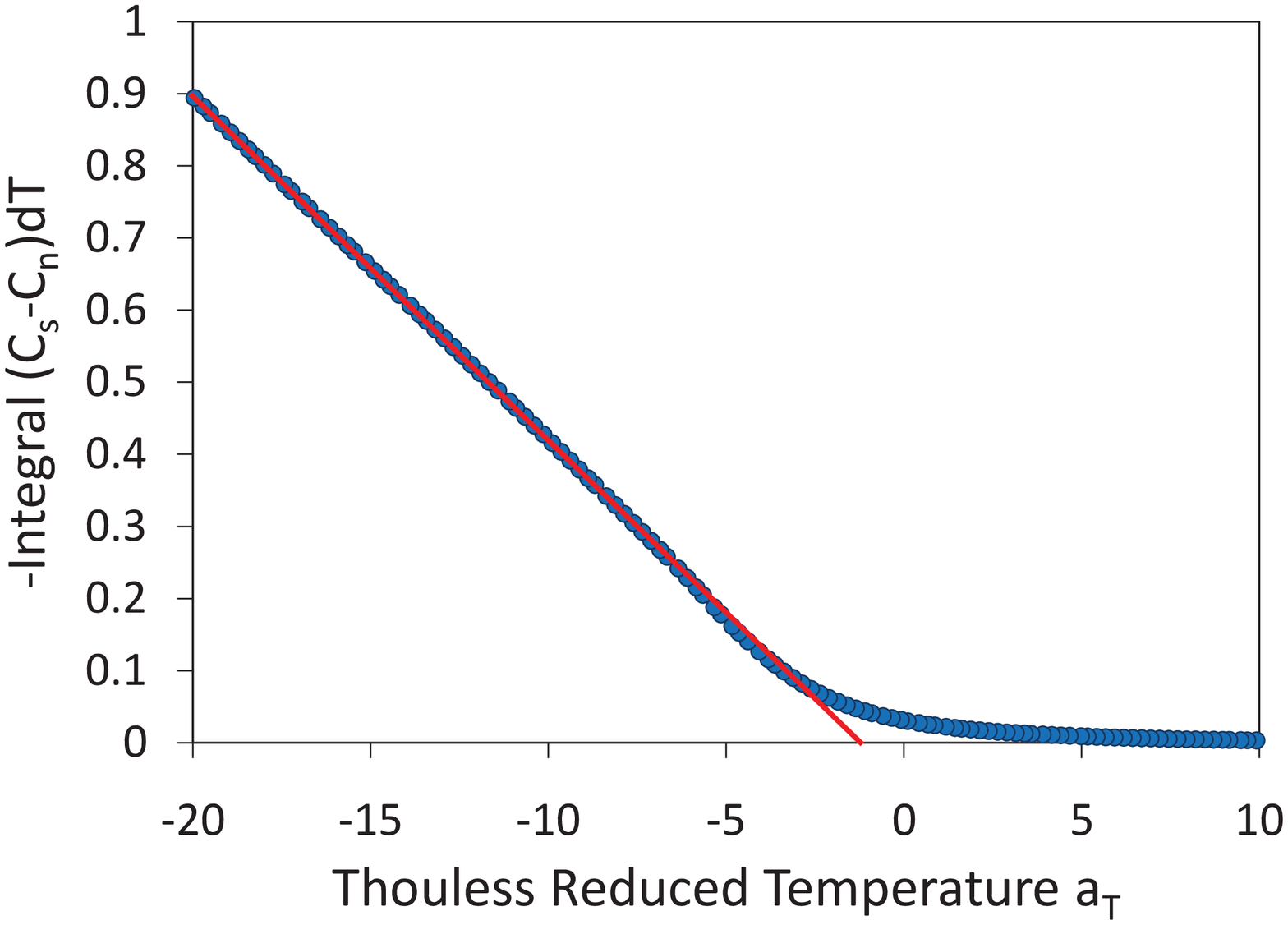}\tabularnewline
\textbf{c} & \includegraphics[bb=50bp 50bp 520bp 700bp,clip,angle=270,width=8.5cm]{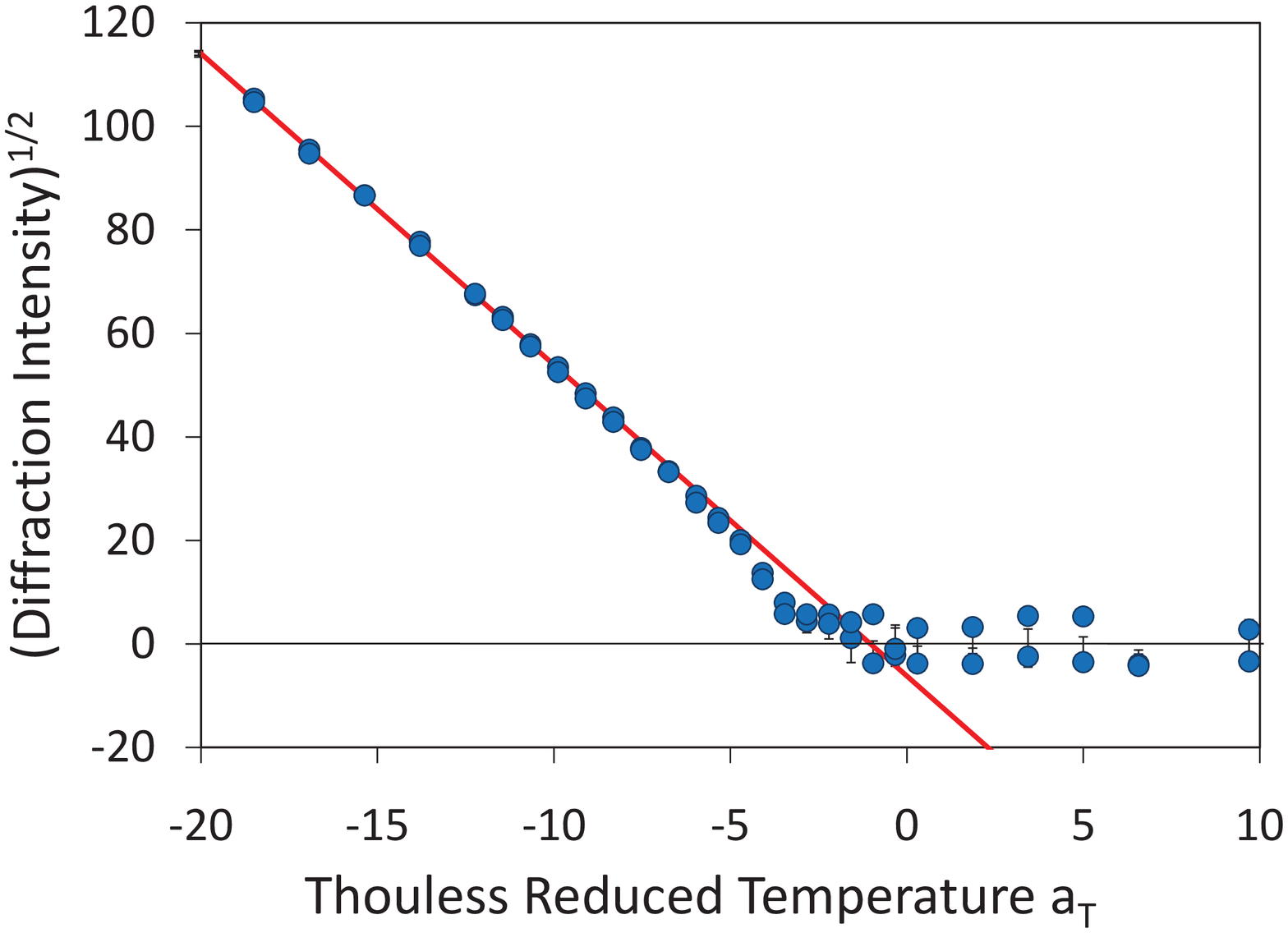}\tabularnewline
\end{tabular}

\caption{(color online) The magnetisation extracted from the heat capacity
and the SANS intensity. (a) \emph{C}$_{s}$ - \emph{C}$_{n}$ from
Thouless theory integrated with respect to temperature plotted versus
the Thouless reduced temperature,\emph{ }a$_{T}$. (b)\emph{ C}$_{s}$
- \emph{C}$_{n}$ from experiment, integrated with respect to temperature
and plotted versus a$_{T}$. Measurements taken from Nb1 in 198 mT.
(c) The SANS diffraction intensity, presented as $\sqrt{I}$ ($\propto$
\emph{M}) versus a$_{T}$. Measurements taken from Nb1 in 200 mT.
Red lines show the extrapolation of linear mean-field-theory behaviour.
\label{fig:aT}}

\end{figure}

As the neutron scattering intensity originates from the magnetisation,
it would be of interest to see how its behaviour compares to that
of the heat capacity. The heat capacity and SANS measurements were
taken using different magnets, and different thermometers, neither
of which had calibration accuracy - in absolute temperature - small
compared with the widths of the transitions. However, by comparison
of the data with theory, we have been able to inter-compare data taken
at different fields and temperatures - and data from the two different
techniques \textendash{} as a function of temperature difference from
the mean field transition temperature, \emph{T}$_{c2,MFT}$, even
though there is no sharp transition once fluctuations are taken into
account. 

To calculate the magnetisation of the sample from the heat capacity,
we use the thermodynamic relation \[
C_{s}-C_{n}=\mu_{0}T\left(\frac{dH_{c2}}{dT}\right)\left(\frac{\partial M_{s}}{\partial T}\right)_{H}.\]

Integrating the Thouless theoretical prediction for the heat capacity
\cite{key-18} down from high temperatures, we obtain Figure \ref{fig:aT}(a),
which indicates that at temperatures below the critical region the
magnetisation varies linearly with temperature, which is the mean
field GL result in the mixed state. However, it will be noticed that
the magnetisation extrapolates linearly to zero at a reduced temperature
of a$_{T}$ = -1.0. This indicates, as one would expect, that fluctuations
not only broaden the transition, but also suppress it below the mean
field value. In Figure \ref{fig:aT}(b), experimental heat capacity
data is integrated in the same way and a similar graph is obtained.
Finally, we consider the SANS data, where below the critical region
the SANS intensity obeys the mean-field result $\sqrt{I}$ $\propto$
\emph{M}. The SANS intensity measurements can be placed on the same
temperature scale by taking the extrapolation of the mean field behaviour,
where $\sqrt{I}$ $\rightarrow$ 0 at a$_{T}$ = -1.0, as shown in
Figure \ref{fig:aT}(c).

We can now compare the temperature-dependence of the magnetisation
obtained from these two different techniques. The magnetisation derived
from the heat capacity measurements demonstrates superconducting fluctuations
result in a long tail above $T_{c2,MFT}$, giving in this region a
diamagnetic magnetisation larger than the mean field extrapolation.
However, the SANS intensity decreases faster than mean field behaviour
and is already immeasurably small below $T_{c2,MFT}$. The different
temperature dependencies is because the heat capacity measurements
are sensitive to all fluctuations that give rise to an average magnetisation,
while the SANS intensity is sensitive only to a spatially coherent
contrast in the magnetisation. Fluctuations of the superconducting
order parameter are expected to result in local field fluctuations
in a low-$\kappa$ superconductor such as niobium \cite{key-18}.
Hence these local field fluctuations reduce the spatially coherent
magnetic field contrast, causing the decrease in SANS intensity below
the mean-field value. This is confirmed in Figure \ref{fig:fluct}(b),
where the SANS intensity in the critical region is shown to obey scaling
once plotted on the a$_{T}$ temperature scale.

\section{Oscillating Field}

\begin{figure*}
\begin{tabular}{cccc}
\textbf{a} & \includegraphics[bb=140bp 50bp 780bp 550bp,clip,width=8.5cm]{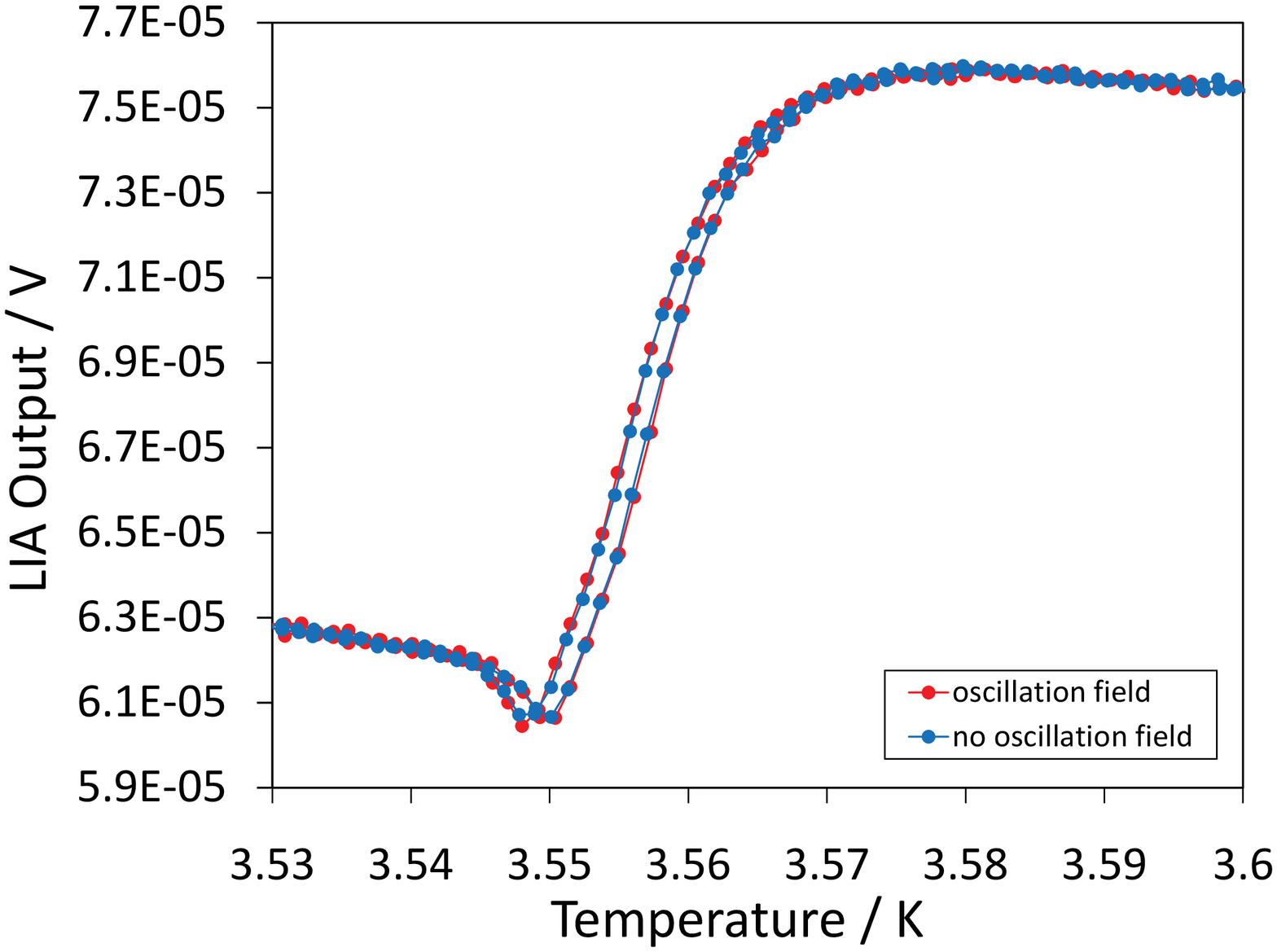} & \textbf{b} & \includegraphics[bb=140bp 50bp 780bp 550bp,clip,width=8.5cm]{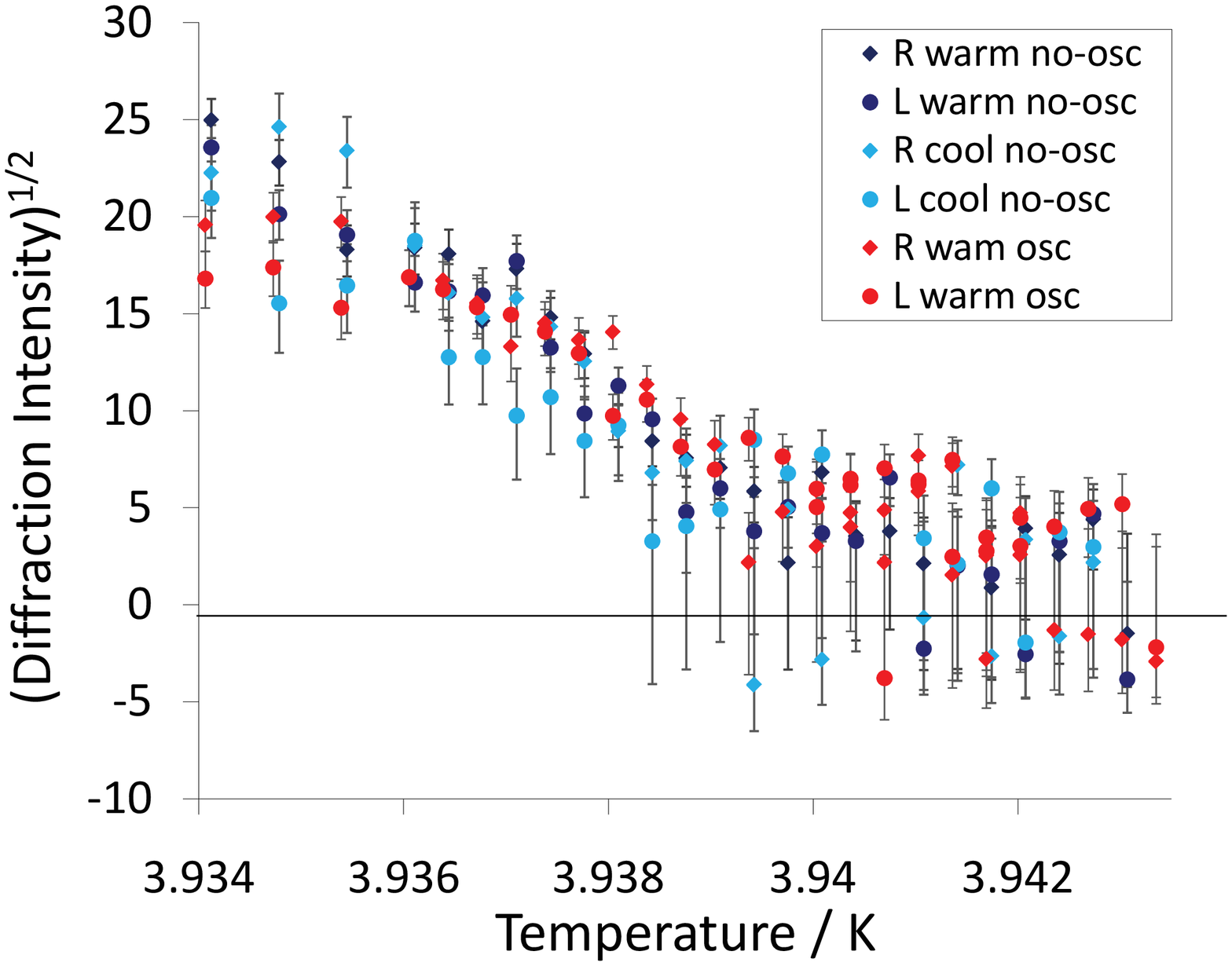}\tabularnewline
\end{tabular}

\caption{(color online). Investigation of a small oscillatory field applied
in addition to the main field. (a) Raw LIA output ($\propto1/C)$
for Nb1 in a field of 308 mT. In one measurement an oscillating field
of amplitude 0.7 G, frequency 100 Hz was continually applied to the
sample during the measurement. (b) Diffracted intensity from Nb3 taken
with and without a small decaying oscillatory magnetic field applied
to the sample before each measurement. The main field applied was
300 mT. \label{fig:wig}}

\end{figure*}

We also performed the heat capacity and SANS measurements with a small
oscillating magnetic field applied to the sample in addition to the
main field. This type of field can help the vortices find their equilibrium
position in the case of weak pinning. Figure \ref{fig:wig}(a) shows
the raw lock-in-amplifier output from the heat capacity measurements
(the signal is $\propto$ $1/C$ and the output has not been corrected
for time lag due to LIA time constant) taken under identical conditions
except for the presence/absence of a small oscillatory magnetic field.
The size of the magnetic field was too small to give a noticeable
change in transition temperature in this scan (and because it was
applied perpendicular, the value of the field was essentially unaltered;
even parallel, $\Delta B_{c2}$ $\sim$ 0.7 G corresponds to $\Delta T_{c2}$
$\sim$ 0.2 mK). The measurements with and without the oscillatory
field are identical. Figure \ref{fig:wig}(b) shows a similar experiment
performed on the SANS measurements, except in this case the oscillatory
field was applied between each measurement, and consisted of a decaying
exponential oscillatory field, with initial amplitude 0.1 G. No difference
in diffracted intensity between these two measurements can be observed
above the noise. That we observed identical behaviour, both with and
without this additional oscillating field, illustrates the very low
pinning nature of our sample and confirms that all our measurements
were performed on a vortex lattice in equilibrium.

\section{CONCLUSIONS}

In summary, measurements of the heat capacity and SANS intensity in
high purity niobium show no sign of a first order vortex lattice melting
transition. Direct observation of the vortices by SANS always finds
a well ordered lattice. Fluctuations are observed to broaden the second
order superconducting to normal transition into a crossover in both
the heat capacity and SANS intensity measurements. Lowest Landau Level
(LLL) scaling is successful over a large range of fields, characterising
the nature of the fluctuations. The fluctuation enhancement of the
heat capacity is found to be more sharply peaked than the predictions
of Thouless\cite{key-18}.

We now have to consider why the vortex lattice does not melt in niobium,
a task made difficult due to the paucity of theoretical work. Our
measurements show that the temperature range over which thermal fluctuations
affect the superconducting order parameter is in agreement with LLL
theory. Calculations within LLL theory predict vortex lattice melting
to occur at a$_{T}$ = -9.5 but in a strongly Type-II superconductor
\cite{key-20}. This is well below our fluctuation region. How this
prediction is modified for low-$\kappa$ niobium is a current theoretical
challenge. One difference is that in the high-$T_{c}$ superconductors
the melting temperature is well below the mean-field $T_{c2}(H)$
so the order parameter is not changing much in magnitude with temperature,
however, in niobium the melting temperature is expected to be very
close to $T_{c2}(H)$ and therefore the loss of phase coherence of
the order parameter would occur in a region where it is rapidly changing. 

The absence of melting may be due to interactions of the vortex lattice
with the underlying crystal lattice, which is known to have a significant
influence on vortex behaviour at lower temperatures. For example,
frustration between the ideal hexagonal vortex lattice and the cubic
crystal lattice results in square, scalene and isosceles vortex lattice
structures when the field is applied along a {[}001{]} crystal axis
of niobium. These structures have been confirmed to persist up to
$H_{c2}(T)$ to within 100 mK of $T_{c2}(H)$ \cite{key-27}. These
strong interactions with the crystal could maintain vortex lattice
rigidity, preventing melting. It would be interesting to expand our
measurements to applied field directions other than parallel to the
{[}111{]} crystal axis, to examine the importance of this interaction
in our fluctuation temperature regime. 

A more unconventional explanation for the absence of vortex lattice
melting in niobium is also a possibility. While there is no doubt
that a first order transition occurs in the high-$T_{c}$ superconductors,
the microscopic details of the transition are still poorly understood,
especially as direct observation of the vortices above the transition
has not been achieved. Our results are compatible with the controversial
proposal \cite{key-21} that the vortex lattice is actually a liquid,
but with a correlation length so large that it is experimentally indistinguishable
from a solid, and the melting transition observed in the high-\emph{T}$_{c}$
superconductors is related to decoupling of pancake layers and would
not be expected to occur in our isotropic system.
\begin{acknowledgments}
This work was supported by the UK EPSRC. Neutron experiments were
performed on instrument D22 at the Institut Laue Langevin, Grenoble,
France. We are grateful for discussions with M. A. Moore, B. Rosenstein
and B. Ya. Shapiro.
\end{acknowledgments}

\section*{Appendix}

Vortex lattice melting in YBa$_{2}$Cu$_{3}$O$_{7-\delta}$ was accompanied
by an entropy change of $\Delta$\emph{S} = 0.4 k$_{B}$ per vortex
per layer\cite{key-2}. We wish to make an estimate of the entropy
change that would accompany vortex lattice melting in niobium. In
comparing the latent heat in a layered system (where the vortex layer
spacing is taken to be the distance between the CuO$_{2}$ planes)
with that in isotropic niobium, we take the layer spacing to be equivalent
to the coherence length, so in niobium we expect $\Delta$\emph{S}
= 0.4 k$_{B}$/vortex/$\xi$, with $\xi$ = 29 nm. (The work of Dodgson
et al. \cite{key-28} suggests that the relevant length scale is the
vortex spacing and not $\xi$; however their theory applies in the
London limit, and in the case considered below there is only a factor
$\sim$ 2 difference between these two lengths). The entropy density,
$\Delta$\emph{s}, is related to the entropy per vortex per layer,
$\Delta$\emph{S}, by $\Delta$\emph{S} = \emph{d}$^{2}$$\xi$$\Delta$\emph{s},
where \emph{d} is the inter-vortex spacing and $\xi$ is the coherence
length/layer spacing. In a field of 300 mT (0.7\emph{H}$_{c2}$(T
= 0), giving \emph{T}$_{c2}$= 3.5 K) the inter-vortex spacing is
\emph{d} = 7.7\texttimes{}10$^{-8}$ m. This gives an entropy density
of $\Delta$\emph{s} = 32 mJ m$^{-3}$ K$^{-3}$ = 3.48 \texttimes{}
10$^{-4}$ mJ mol$^{-1}$ K$^{-1}$.

We can make an estimate for the height of the heat capacity spike
by assuming the spike is triangular in shape, with a transition width
of $\Delta$\emph{T}$_{melt}$. The width of this hypothetical melting
transition will be related to sample homogeneity. In zero field (where
fluctuations are only expected to be present on the scale of nK) our
\emph{T}$_{c2}$(0) width of 0.4 mK (measured from the heat capacity)
is testament to the high quality of the sample. In a magnetic field
we need to consider how much any \emph{variations} in sample purity
will broaden \emph{T}$_{c2}$, (additional to any broadening due to
fluctuations of the order parameter). The upper critical field, \emph{B}$_{c2}$,
depends on the Ginzburg-Landau parameter, $\kappa_{1}$, as \emph{B}$_{c2}$(T)
= $\sqrt{2}$$\kappa_{1}$\emph{B}$_{c}$(T), and $\kappa_{1}$ depends
on the electronic mean free path as given \cite{key-24} by $\kappa$
= $\kappa_{0}$ + 7500$\rho_{0}$$\sqrt{\gamma}$, where $\kappa_{0}$
is for the case of infinite mean free path, $\rho_{0}$ is the residual
resistivity and $\gamma$ is the Sommerfeld constant. Any variations
in sample purity will give a spread in residual resistivity across
the sample, resulting in a spread in upper critical field, $\Delta$\emph{B}$_{c2}$
$\sim$ ($\xi$$\Delta$\emph{l}/\emph{l}$^{2}$)\emph{B}$_{c2}$,
where \emph{l} is the electron mean free path. This estimation emphasises
the importance of using a high-purity sample.

We can relate \cite{key-25} electron mean free path to the residual
resistivity ratio, \emph{R}, using \emph{l} = 5 \texttimes{} 10$^{-9}$\emph{R},
so for our sample with \emph{R} = 1000, \emph{l} = 200$\xi$. Therefore,
using $\Delta$\emph{T}$_{c2}$ $\sim$ ($\xi$$\Delta$\emph{l}/\emph{l}$_{2}$)\emph{B}$_{c2}$(d\emph{B}$_{c2}$/d\emph{T})$^{-1}$,
we have, for an assumed inhomogeneity in\emph{ $\Delta$l} of 5 \%,
a transition width due to inhomogeneity broadening of $\Delta$\emph{T}$_{c2}$
$\sim$ 2 mK in a field of 300 mT. If the width of the melting transition
is the same: $\Delta$\emph{T}$_{melt}$ $\sim$ 2 mK, we can therefore
estimate the height of the spike in the heat capacity to be $\Delta$\emph{C}$_{melt}$
$\sim$ 1.2 mJ mol$^{-1}$ K$^{-1}$ in a field of 300 mT. The size
of the mean field heat capacity jump in this field is $\Delta$\emph{C}$_{MFT}$
= 10 mJ mol$^{-1}$ K$^{-1}$ so the spike would be 12 \% of the jump.

\end{document}